\documentclass[12pt]{article}
\usepackage{amssymb}
\def\inh{\vskip 0.075truein \noindent\hangindent=12 pt \hangafter=1}

\usepackage{amsthm}\theoremstyle{remark}

\newcommand{\bte}{\begin{quote}\begin{theorem}}
\newcommand{\ete}[1]{\label{#1}\end{theorem}\end{quote}}
\newcommand{\bcom}{\begin{quote}\begin{comment}}
\newcommand{\ecom}[1]{\label{#1}\end{comment}\end{quote}}
\newcommand{\bex}{\begin{quote}\begin{example}}
\newcommand{\eex}[1]{\label{#1}\end{example}\end{quote}}
\newcommand{\bcon}{\begin{quote}\begin{conclusion}}
\newcommand{\econ}[1]{\label{#1}\end{conclusion}\end{quote}}
\newcommand{\bdefi}{\begin{quote}\begin{definition}}
\newcommand{\edefi}[1]{\label{#1}\end{definition}\end{quote}}

\newcommand{\blem}{\begin{quote}\begin{lemma}}
\newcommand{\elem}[1]{\label{#1}\end{lemma}\end{quote}}

\newcommand{\bpr}{\begin{quote}\begin{problem}}
\newcommand{\epr}[1]{\label{#1}\end{problem}\end{quote}}

\newcommand{\f}{\frac}
\newcommand{\p}{\partial}
\newcommand{\n}{\nonumber \\}

\newcommand{\beq}{\begin{eqnarray}}
\newcommand{\eeq}[1]{\label{#1}\end{eqnarray}}
\newcommand\eq[1]{(\ref{#1})}
\newcommand{\bfi}{\begin{figure}[24]}
\newcommand{\efi}[1]{\caption{\label{#1}}\end{figure}}
\newcommand\fig[1]{Fig.~\ref{#1}}
\newcommand{\res}{respectively}
\newcommand\gl{\left}
\newcommand\gr{\right}
\newcommand{\Ga}{\alpha}
\newcommand{\Gb}{\beta}
\newcommand{\Gd}{\delta}

\newcommand{\Gf}{\phi}

\newcommand{\Gk}{\varkappa}

\newcommand{\Gn}{\eta}

\newcommand{\Gr}{\varrho}

\newcommand{\Go}{\omega}

\newcommand{\GF}{\Phi}

\newcommand{\GO}{\Omega}

\newcommand\D{\,\mathrm{d}}
\newcommand\I{\mathrm{i}}
\newcommand\E{\mathrm{e}}
\newcommand{\bexe}{\begin{quote}\begin{exercise}\inh}
\newcommand{\eexe}[1]{\label{#1}\end{exercise}\end{quote}}

\usepackage{graphics,graphicx}
\usepackage{color}
\begin{document}

{\large
\title{Forerunning mode transition in a continuous waveguide}
}

\author{Leonid Slepyan$^{a*}$, Mark Ayzenberg-Stepanenko$^{b}$, Gennady Mishuris$^{c}$}
\date{\small{$^a${\em School of Mechanical Engineering, Tel Aviv University\\
P.O. Box 39040, Ramat Aviv 69978 Tel Aviv, Israel}\\
$^b${\em The Shamoon College of Engineering, Beer-Sheva 84105, Israel} \\
 $^c${\em Institute of Mathematics and Physics, Aberystwyth University\\
Ceredigion SY23 3BZ Wales UK} \\
}}

\maketitle

\vspace{10mm}\noindent
{\bf Abstract}
We have discovered a new, forerunning mode transition as the periodic transition wave propagating in a uniform continuous waveguide. The latter is represented by an elastic beam separating from the elastic foundation under the action of sinusoidal waves.
The critical displacement is the separation criterion. We show that the steady-state separation mode, where the separation front speed is independent of the wave amplitude, exists only in a bounded speed-dependent range of the wave amplitude.  As the latter exceeds the critical value the steady-state mode is replaced by a more complicated regime with porous-like local separations emerging at a distance ahead of the main transition front. The crucial feature of this simple model is that the wave group speed is greater than the phase speed. The analytical solution allows us to show in detail how the steady-state mode transforms into the forerunning one. The established forerunning mode studied numerically manifests itself as a periodic process. As the incident wave amplitude grows the period decreases, while the separation front speed averaged over the period increases to the group velocity of the wave. In addition, the complete set of relations is presented for the waves excited by the oscillating force moving along the free beam, including the resonant case corresponding to a certain relation between the load's speed and frequency.

\vspace*{10mm}\noindent

\vspace{10mm}\noindent
$^*$ Corresponding author (Leonid Slepyan), email: leonid@eng.tau.ac.il
\section{Introduction}
There exist processes like dynamic crack growth, phase transition or other similar events, where a change of the body structure or state spreads as a wave. The {\em transition wave} can propagate being forced by the action of external loads or spontaneously drawing energy initially distributed in the waveguide (Ayzenberg-Stepanenko et al, 2014). Formulation of such problems includes equations for both states and a transition criterion or equations for the transition. In accordance with the constitutive equations, the transition can occur instantaneously at the wave front or during a period. Note that the former mode of transition is much simpler for mathematical analysis. In this case, the waveguide is separated by a moving transition point (or an interface) into two parts, the intact part is placed in front of this point, while the modified part or a different state appears behind this point. Note that generally, in the framework of a continuous material, the formulation of the transition criterion is not trivial. This question is discussed in detail in Slepyan (2002).

In analytical studies, transition waves are commonly considered under the steady-state formulation assuming that the dynamic state depends on variable $\Gn=x-Vt$ but not on $x$ and $t$ separately ($x$ is the continuous or discrete coordinate, $t$ is time and $V$ is the transition wave speed). Naturally, the wave speed is defined by the type and level of the action; however, an inverse problem is usually studied, where the speed is an input parameter. In this case, the corresponding dynamic equations, conditions at infinity and the transition criterion posted at a point of the $\Gn$-axis, say, $\Gn=0$, uniquely define the steady-state solution.

However, in addition, such a solution must satisfy the admissibility condition (Marder and Gross, 1995), which states, in general, that the transition criterion should not be satisfied before the moment assumed in the problem formulation, that is, it should not be satisfied at $\Gn>0$ or outside the considered line. Note that this condition was stated for the lattice fracture; however, it is valid for the problem under consideration as well. In many cases, this condition essentially bounds the region of existence of the steady-state solutions.

It was observed in the first analytical work on the lattice fracture (Slepyan, 1981a) that there are different crack speed solutions corresponding to a given `macrolevel' energy release rate. For this case, in the plane problem, the Marder-Gross condition results in the conclusion that only the highest crack speed steady-state solution is admissible if no fracture occurs outside of the prospective crack line.

 We note in this respect that there is an essential difference in steady-state transition waves excited by constant unmoving and moving-oscillating loads. In the former case, where the incident wave is infinitely long, the speed of the transition point increases approaching the incident wave group velocity as the force increases. In contrast, in the latter case, where the incident wave is of a finite length, the steady-state transition can exist if the group speed exceeds the phase speed, that is, in the case of the so called `anomalous dispersion'. In this instance, the transition speed equals to the latter independently of the load level, and only the transition point position relatively to the incident wave depends on the load.

The analytical solutions for fracture under a sinusoidal wave was presented by Slepyan (1981b, 2010) for a lattice and a continuous body, \res. The papers most related to the considered issue are those for the fracture dynamics in the lattice waveguide, Mishuris et al (2009) and Slepyan et al (2010). It was found in the former and explained in more detail in the latter that there exists a piecewise constant relationship between the crack speed and the wave amplitude under a fixed wave frequency. In a bounded wave amplitude region, the crack speed is equal to the wave phase speed, and the steady-state solution is valid. Then, as the wave amplitude exceeds the critical level, in the next wave amplitude region, the two-bond clustering occurs with two alternating values of the crack speed, and the averaged crack speed again is constant but greater than the phase speed of the wave. In the further increase of the wave amplitude, the number of the bonds in the cluster increases, while the averaged crack speed is constant in each corresponding wave amplitude region. As the wave amplitude grows this averaged-over-the-cluster crack speed approaches the group speed of the wave, as it should be. Such a clustering was also observed in the spontaneous crack propagation in a a two-line chain with internal potential energy, Ayzenberg-Stepanenko et al (2014). Note that transition waves in lattices were considered in many works, see, e.g., Slepyan and Ayzenberg-Stepanenko (2004) and Vainchtein (2010) and the references herein.

In the present work, we consider a beam on an elastic foundation with the aim to find what the transition modes can exist in the continuous waveguide under the action of sinusoidal waves.
In this simple model, the transformation of the steady-state mode into more complicated, forerunning transition mode can be observed in detail. The established forerunning mode manifests itself as a periodic process. As the incident wave amplitude grows the period decreases, while the separation front speed averaged over the period increases to the group velocity of the wave. In contrast to the discrete structures, both these parameters continuously depend on the wave amplitude.

Note that transition waves in this model, including some different versions of it, was previously studied by Brun et al (2013), where the transition to a lower nonzero stiffness of the foundation was considered propagating under the action of gravity. In particular, the steady-state solution was presented for the intersonic transition wave speed, where it could exist only.

The paper is organized as follows. In the next auxiliary section, we present the complete set of relations for the waves excited by the moving-oscillating force in the free beam. The relations between the amplitude, speed and frequency of the load and the wave parameters are presented including those for the resonant case corresponding to a certain relation between the load's speed and frequency. Note that numerous problems related to the moving load are considered in the book by Friba (1999), see also Cai et al (1988).
Next, the analytical solution for the steady-state transition under the sinusoidal wave is presented, and the bounds of the domain in the wave speed-amplitude plane are found, where the steady-state mode exists. Further, the results of numerical simulation of the forerunning transition mode are given and discussed.

\section{Flexural waves excited by a moving-oscillating force in a free beam}
\subsection{The equation, dispersion relations and waves}
We start with the Bernoulli-Euler equation for a beam
 \beq EI \f{\p^4 w(x,t)}{\p x^4} + \Gr A \f{\p^2 w(x,t)}{\p t^2} = P\Gd(x-vt)\E^{\I\Go t}\,,\eeq{1}
where $EI, A$ and $\Gr$ are the bending stiffness, cross-section area and density of the beam, $w$ is its transverse displacement, $x$ and $t$ are the coordinate and time and $P$ is the amplitude of the force moving along the beam with constant speed $v$ and oscillating with frequency $\Go$.

We take $r=\sqrt{I/A}$ and $c=\sqrt{E/\Gr}$ as natural length and speed units (accordingly, $r/c$ is the time unit), and $EA$ the force unit. In terms of the corresponding non-dimensional variables (we use the same notations), the equation becomes
\beq \f{\p^4 w(x,t)}{\p x^4} + \f{\p^2 w(x,t)}{\p t^2} = P\Gd(x-vt)\E^{\I\Go t}\,.\eeq{2}

The force excites sinusoidal waves, which frequencies, $\GO$, and wave numbers, $k$, are defined by the dispersion and Doppler relations
 \beq \GO=\pm k^2\,,~~~\GO=\Go+k v\,.\eeq{dr1}
It follows from this that
 \beq k_{1,2}=-\f{1}{2}\gl(v\pm \sqrt{v^2-4\Go}\gr)~~(v^2\ge 4\Go)\,,~~~k_{1,2}=-\f{1}{2}\gl(v\pm \I\sqrt{4\Go-v^2}\gr)~~(v^2\le 4\Go)\,,\n
k_{3,4}=\f{1}{2}\gl(v\mp \sqrt{v^2+4\Go}\gr)\,,\eeq{dr2}
where the nonzero real $k$ correspond to the sinusoidal waves with the phase and group speeds
 \beq V=\f{\GO}{k} =\sqrt{|\GO|}\mbox{sign}\,(\GO k)\,,~~~V_g=2V\,.\eeq{fgs1}
The wave propagates in front of the moving-oscillating force, at $\Gn>0$, if its group velocity exceeds $v$; otherwise, it is placed at $\Gn<0$.

\begin{figure}[h!]
\centering
    \includegraphics [scale=1.15]{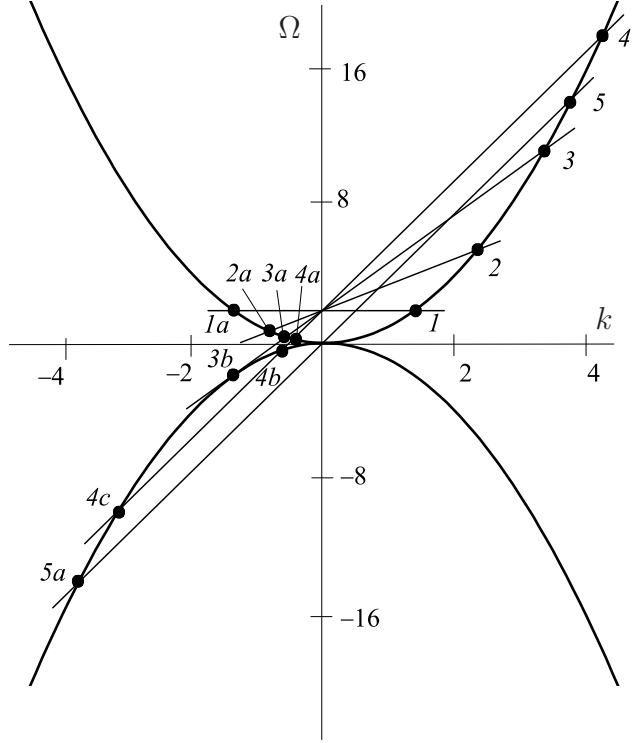}
\begin{picture}(0,0)(0,-200)
\put(-160,80){$\Omega$}
\put(-40,-30){$k$}
\end{picture}

\caption{Dispersion diagram for the free beam, $\GO=\pm k^2$, and the Doppler rays, which intersections with the dispersion curves, marked by 1, 1a, ... 5, 5a, correspond to the wave numbers, $k$, and frequencies, $\GO$, of the sinusoidal waves excited by the force moving with speed $v$ and oscillating with frequency $\Go$. The rays correspond to: an unmoving force, $\GO=\Go>0, v=0$ (1-1a), to a subcritical speed, $\Go>0, 0<v<v_c$ (2-2a), to the resonant regime, $\Go>0, v=v_c=2\sqrt{\Go}$ (3-3a-3b), to a supercritical speed, $\Go>0, v>v_c$ (4-4a-4b-4c) and to a non-oscillating moving force, $\Go=0, v>0$ (5-5a). The latter ray has also an intersection with both dispersion curves at the origin, $\GO=k=0$.  }
\label{f1}
\end{figure}

The dispersion diagram and the Doppler rays \eq{dr1} are shown in \fig{f1} for different speeds for a value of $\Go>0$ and for $\Go=0, v>0.$

\subsection{Wave amplitudes}
We consider the steady-state solution as the limit of the transient solution corresponding to zero initial conditions. We derive it directly starting from the Laplace and Fourier transforms on $t$ and $x$ \res. It follows from \eq{2} that
\beq w^{LF}(k,s)=\f{P}{[s-\I(\Go+kv)](k^4+s^2)}\,,\n
 w(x,t) = \f{1}{4\pi^2 \I}\int_{-\I\infty +0}^{\I\infty+0}\int_{-\infty}^\infty \f{P\E^{-\I k x +s t}}{[s-\I(\Go+kv)](k^4+s^2)}\,.\eeq{lf1}
Substituting $x=vt+\Gn, s=s'+\I(\Go+kv)$ we obtain
\beq w^{LF_\Gn}(k,s') = \E^{\I\Go t}W^{LF_\Gn}(k,s')\,,~~~W^{LF_\Gn}(k,s')=\f{P}{s'[k^4 +(s'+\I(\Go+kv))^2]}\,,\eeq{lf2}
where the superscript $F_\Gn$ denotes the Fourier transform on $\Gn$. Assuming that the limit exists we find
 \beq W^{F_\Gn}(k) = \lim_{s'\to +0} s'W^{LF_\Gn}(k,s') = \f{P}{k^4 -(\Go+kv -\I 0)^2]}\,.\eeq{lf3}

A complete steady-state solution follows from \eq{lf3} as a sum of four residues in the Fourier inverse transform. The waves propagating to the right are defined by the real zeros of the denominator in \eq{lf3} coming to the real axis from below ($k=k_i-0$), and vice versa.  it follows that for $v < 2\sqrt{\Go}$ the wave amplitude is
 \beq
W(\Gn) &=& -\f{1}{4}\gl(\f{\I\E^{-\I k_4 \Gn}}{\GO_4\sqrt{\Go+v^2/4}}
+\f{\E^{-\I k_1\Gn}}{\GO_1\sqrt{\Go-v^2/4}}\gr)PH(\Gn)\,,\n
W(\Gn) &=& -\f{1}{4}\gl(\f{\I\E^{-\I k_3 \Gn}}{\GO_3\sqrt{\Go+v^2/4}}
+\f{\E^{-\I k_2\Gn}}{\GO_2\sqrt{\Go-v^2/4}}\gr)PH(-\Gn)\,,
 \eeq{a1}
where
\beq \GO_i= \Go +k_iv\,,~~~k_{1,2}=-\f{1}{2}\gl(v\pm\I\sqrt{\Go-v^2/4}\gr)\,~~~k_{3,4}=\f{1}{2}\gl(v \mp\sqrt{\Go+v^2/4}\gr)\,.\eeq{a2}

There is no steady-state in the resonant excitation, $v=2\sqrt{\Go}$. The transient problem is considered separately. For $v>2\sqrt{\Go}$ two waves propagate in each direction.
 \beq W(\Gn) &=& -\f{\I}{4}\gl(\f{\E^{-\I k_4 \Gn}}{\GO_4\sqrt{\Go+v^2/4}} + \f{\E^{-\I k_1 \Gn}}{\GO_1\sqrt{v^2/4-\Go}}\gr)PH(\Gn)\,,\n
W(\Gn)& = &-\f{\I}{4}\gl(\f{\E^{-\I k_3 \Gn}}{\GO_3\sqrt{\Go+v^2/4}} + \f{\E^{-\I k_2 \Gn}}{\GO_2\sqrt{v^2/4-\Go}}\gr)PH(-\Gn)\,,\eeq{a3}
where
\beq \GO_i= \Go +k_iv\,,~~~k_{1,2}=-\f{v}{2}\mp \sqrt{v^2/4-\Go}\,~~~k_{3,4}=\f{v}{2}\mp\sqrt{\Go+v^2/4}\,.\eeq{a4}

In a special case on a non-oscillating moving force, $\Go=0, v>0$, referring to \eq{lf3} we have
 \beq W^{F_\Gn}(k) =  \f{P}{(k-v+\I 0)(k+v+\I 0)(k+\I 0)^2}\,.\eeq{lf3a}
It follows from here that
 \beq W(\Gn)= -\f{P\sin v\Gn}{v^3}H(\Gn) -\f{P\Gn}{v^2}H(-\Gn)\,.\eeq{a5}
Note that this result also follows from \eq{a3} as the limit at $\Go=+0$.

In a particular case of an unmoving oscillation force
\beq k_1=-k_2=\I\sqrt{\Go}\,,~~~k_4=-k_3=\sqrt{\Go}\,,~~~\GO_{3,4}=\Go\,,\eeq{a6}
and it follows from \eq{a1} that
 \beq
W(\Gn) = -\f{P}{4\Go^{3/2}}\gl(\I\E^{-\I \sqrt{\Go} |\Gn|}+\E^{- \sqrt{\Go}|\Gn|}\gr)\,.\eeq{a7}

Note that in the transition to dimensional quantities, one should make replacements in accordance with the above definitions, namely
 \beq W \to \f{W}{r}\,,~~~(\GO,\Go) \to (\GO,\Go)\f{r}{c}\,,~~~v \to \f{v}{c}\,,~~~P \to \f{P}{EA}\,.\eeq{ndvr1}

Thus, the steady-state solutions exist for any values of the speed and frequency except cases $v=\Go=0$ and $v=2\sqrt{\Go}$. In addition to this, there exist transient regimes as weakly localised growing oscillations. These regimes having no steady-state limit are considered below.

\subsection{Resonant waves}

Clearly, the solutions in \eq{a1}, \eq{a3} fail for $v=2\sqrt{\Go}$. For this special case, where the wave group velocity coincides with the load velocity (see \fig{f1}), $k_1=k_2=-v/2$, and we put in \eq{lf2} $k=q-v/2$. As a result, we have
 \beq W^{LF_\Gn}(s,k) = \f{P}{s'[(q-v/2)^4 +(s'+\I(vq-v^2/4))^2]}\n
  = \f{P}{s'(s'+\I q^2)[s'-\I(q^2-2vq+v^2/2)]}\,.\eeq{rw1}
Only the poles at $s=0$ and $s'=-\I q^2$ should be taken into account to obtain an asymptote for the growing resonant wave, and the latter is given by the integration, in the inverse Fourier transform, over an arbitrary small vicinity of the point $q=0 \, (k=-v/2)$. It follows that for $\Go>0$
 \beq w(x,t) \sim  \f{P}{\pi \Go}\E^{\I (\GO t-kx)}\GF_2(\xi)\,,~~~\xi=\f{x}{2\sqrt{t}}\,,~~\GO=-k^2\,,~~k=-\f{v}{2}\,,\n
 \GF_2=\int_0^\infty\f{2\sin^2(q^2t/2)+\I\sin (q^2t)}{q^2}\cos(q\Gn)\D q\n
 = \f{1}{\sqrt{\pi}}\E^{\I(\xi^2+\pi/4)} +
 \xi\gl[\mbox{FresnelS}(\xi\sqrt{2/\pi}) +\mbox{FresnelC}(\xi\sqrt{2/\pi})\gr]\n
   +\I\xi\gl[\mbox{FresnelS}(\xi\sqrt{2/\pi})-\mbox{FresnelC}(\xi\sqrt{2/\pi})\gr]- |\xi| \eeq{rw2}
with
\beq \mbox{FresnelC}(x) =\int_0^x \cos\gl(\f{\pi}{2}t^2\gr)\D t\,,~~~\mbox{FresnelS}(x) =\int_0^x \sin\gl(\f{\pi}{2}t^2\gr)\D t\,.\eeq{FrenCS}
Note that in the case of a constant unmoving force, $v=\Go=0$, where all the poles matter, the displacement grows as $t^{3/2}$.

\section{Transition wave}
\subsection{Formulation}
Consider the infinite beam initially resting on a massless elastic foundation in the region, $x>0$. This also can be envisioned that the beam is attached to a substrate through a thin elastic layer of glue or something like, that results in the same mathematical formulation. Under a sinusoidal wave, which can be excited by an oscillating or/and moving force as discussed above, the connection to the base breaks at any point where the beam displacement reaches the critical value. So, the foundation remains intact while
 \beq w(x,t) <w_c\,,~~~t<t_*(x)\,,\eeq{kr1}
and disappears at the moment, $t=t_*(x)$, when the displacement reaches the critical value.

The equation for the intact region is
\beq EI \f{\p^4 w(x,t)}{\p x^4} + \Gr A \f{\p^2 w(x,t)}{\p t^2} +\Gk w(x,t)=0\,,\eeq{tw2}
whereas the foundation stiffness $\Gk=0$ in the free beam domain.

In addition, we have five conditions at the separation point: four continuity conditions with respect to
$w(x,t)$, $w'(x,t)$, $w''(x,t)$, $w'''(x,t)$, and the transition criterion, $w(x,t)=w_c$. Besides, the conditions at plus/minus infinity are assumed as follows. A sinusoidal wave of the amplitude $A$ and frequency $\GO$ propagates to the right along the free beam with the phase speed $V=\GO/k$, and there is no other energy flux from plus/minus infinity.

\begin{figure}[h!]
   \centering
   \includegraphics [scale=1.3]{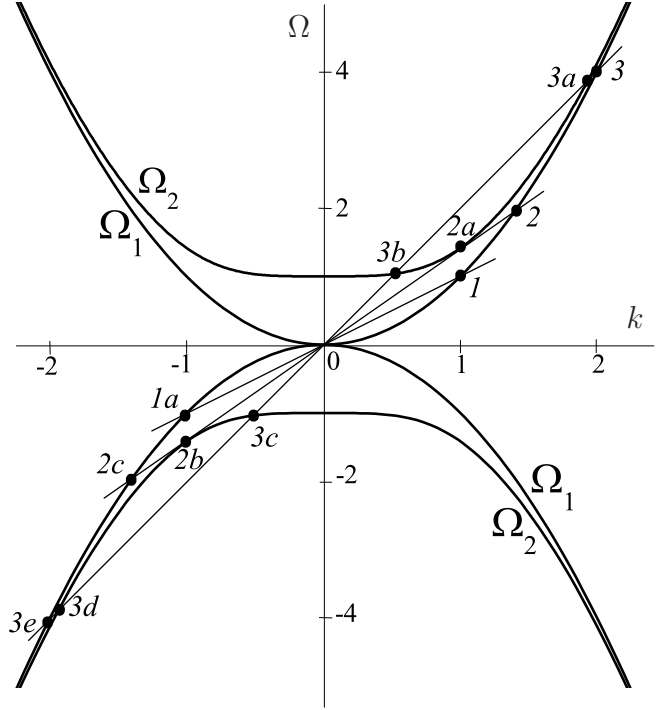}
   \begin{picture}(0,0)(0,-200)
\put(-163,70){$\Omega$}
\put(-35,-40){$k$}
\end{picture}

\caption{Dispersion diagram for the free and supported beams, $\GO_1=\pm k^2$, and $\GO_2=\pm\sqrt{1+k^4}$, \res, and the rays corresponding to the intersonic  (1 - 1a),  supersonic (3 - 3a -3b - 3c - 3d - 3e) wave phase speeds and the separating ray (2 - 2a -2b - 2c) with $V=\sqrt{2}$. The intersections with the dispersion curves, marked by 1, 1a, ... 3d, 3e, correspond to the wave numbers, $k$, and frequencies, $\GO$, of the sinusoidal waves, which are radiated, in the steady-state regime, in front of the transition point, $\Gn>0$, for $\GO=\GO_2$ and $V_{g2}=\D\GO_2/\D k>V_2=\GO_2/k$, that corresponds to the points 2a, 2b and 3a, 3d. It can be seen that no such waves are radiated behind the transition point, because the corresponding condition, $V_{g1}=\D\GO_1/\D k<V_1=\GO_1/k$ does not exist for $\GO=\GO_1$. However, there exists a displacement linearly distributed at  $\Gn<0$ corresponding to intersection point $\GO_1=k=0$.}
\label{f2}
\end{figure}

We use here a different normalisation, namely, we consider the below quantities
 \beq l=\gl(\f{r^2EA}{\Gk}\gr)^{1/4}\,,~~~\tau =\sqrt{\f{\Gr A}{\Gk}}\,,~~~p= \Gk l^2\eeq{nndu1}
as the natural units of length, time and force, \res.
In these terms, the equations become
\beq \f{\p^4 w(x,t)}{\p x^4} + \f{\p^2 w(x,t)}{\p t^2} +w(x,t)=0~~~(\mbox{the intact region})\eeq{fbne1}
and
\beq \f{\p^4 w(x,t)}{\p x^4} + \f{\p^2 w(x,t)}{\p t^2} =0~~~(\mbox{the free beam region})\,.\eeq{sbne1}
The dispersion relations corresponding to these equations are
 \beq \GO=\GO_1=\pm k^2~~~(\mbox{the free beam})\,,~~~\GO=\GO_2=\pm \sqrt{1+k^4}~~~(\mbox{the supported beam})\,\eeq{dr1a}
as shown in \fig{f2}.

\vspace{3mm}

\subsection{Steady-state regime and the domains of its existence}
Consider the steady-state regime, which exists only for intersonic speeds (as already was noted in Brun et al (2013)). Eqs. \eq{fbne1} and \eq{sbne1} are valid at $\Gn<0$ and $\Gn>0$, \res, and $w=w(\Gn)$, where $\Gn=x-Vt$. Note that the steady state implies that the speed of the separation point coincides with the phase speed of the incident wave, $V$.
The equations become
 \beq \f{\D^4 w(\Gn)}{\D \Gn^4} +V^2 \f{\D^2 w(\Gn)}{\D \Gn^2} +w(x,t)=0~~~(\Gn>0)\,,\n
 \f{\D^4 w(\Gn)}{\D \Gn^4} +V^2 \f{\D^2 w(\Gn)}{\D \Gn^2} =0~~~(\Gn<0)\,.\eeq{sseom1}
In accordance with the dispersion diagram, \fig{f2}, the general solution to these equations contains six unknown constants, two for waves at $\Gn\ge 0$ exponentially decreasing or propagating to the right and four for the waves at $\Gn\le 0$ propagating in both directions. Recall that these constants are defined by the continuity conditions and the transition criterion
 \beq w(-0)=w(+0)=w_c,~~w'(-0)=w'(+0),~~w''(-0)=w''(+0),~~w'''(-0)=w'''(+0),\eeq{ccfssr1}
and the relation
 \beq w_i(-0)+w_r(-0)=w(-0)\eeq{ccftiw1}
for the sum of the incident and reflected waves, \res.  In addition, we choose the steady-state solution satisfying a partial condition of admissibility
\beq \f{\p w(\Gn)}{\p t} =- V\f{\D w(\Gn)}{\D \Gn}>0~~~(\Gn=0)\,.\eeq{pcoa1}

These conditions allow us to construct a uniquely defined steady-state solution (presented below) formally valid for any large incident wave amplitude. The main questions are whether the solution really exist and what mode is formed instead otherwise. To answer the first question we have to check if the Marder-Gross admissibility condition is satisfied. Namely, if the critical displacement is not reached earlier than it is assumed in the problem formulation, that is, if $w(\Gn)<w_c$ for any $\Gn>0$. Below we show that this condition is satisfied and hence the solution is valid only in a bounder domain in the incident wave speed - amplitude plane.  For the incident wave parameters outside of this domain the forerunning transition wave mode is disclosed numerically. We here call intersonic and supersonic regimes for $0<V<\sqrt{2}$ and $V> \sqrt{2}$, \res. Note that the speed $V=\sqrt{2}$ coincides with the group speed, $\D\GO_2/\D k$, of the wave in the intact area. It follows that no sinusoidal wave can propagate to the right in the latter area at this speed, as it is in the intersonic regime.

As for the the supersonic incident wave, it can be concluded in advance that in this regime, no steady-state solution exists. This follows directly from the fact that in such a regime, if it were exist, a sinusoidal wave would propagate at  $\Gn>0$ with the amplitude $w_{max}\ge w_c$ ($w(+0)=w_c$) and thus would violate the Gross-Marder condition.

The solutions at the left and at the right, where some of the conditions in \eq{ccfssr1} and \eq{ccftiw1} are already taken into account, are
 \beq w(\Gn) = A[\cos (V\Gn+\Gf) -\cos\Gf] +w_c +C_1\Gn~~~(\Gn\le 0)\,.\eeq{ssds2}
and
 \beq w(\Gn)=\E^{-\Ga \Gn}[w_c\cos(\Gb\Gn)+C_2\sin(\Gb\Gn)]~~~(\Gn\ge 0)\,,\eeq{ssds4}
 where $C_{1,2}$ are arbitrary constants and
  \beq \Ga = \f{1}{2}\sqrt{2-V^2}\,,~~~\Gb= \f{1}{2}\sqrt{2+V^2}\,.\eeq{ssds5}
The rest continuity conditions concerning the derivatives of the displacement up to the third order at $\Gn=0$ and the partial admissibility condition \eq{pcoa1} define the constants $C_{1,2}$ and the phase shift, $\Gf$. We find
 \beq C_1=\f{AV^2\cos\Gf+(1-V^2)w_c}{2V^2\Ga}\,,~~~ C_2 = \f{V^2[2A\cos\Gf-w_c]}{4\Ga\Gb}\,,\n
 \cos\Gf = - 2V\Ga\sqrt{1-\gl(\f{w_c}{AV^2}\gr)^2}-(1-V^2)\f{w_c}{AV^2}\,.\eeq{ssds6}

It follows from this solution that  the steady-state regime of the transition may exist only if the incident wave amplitude is large enough, namely, if
 \beq A\ge \f{w_c}{V^2}\,.\eeq{tlbfsss}
This can be seen in \eq{ssds6}.
We, however, must check whether the steady-state solution obtained here satisfies the Marder-Gross condition, $w(\Gn) <w_c~ (\Gn>0)$. As can be seen in \eq{ssds4} the first maximum of $w(\Gn)$ at $\Gn>0$ is the global maximum in the intact region, and the admissibility condition is satisfied if it is below $w_c$.

\begin{figure}[h!]

\vspace{-40mm}
    \hspace{-17mm}\includegraphics [scale=2.05]{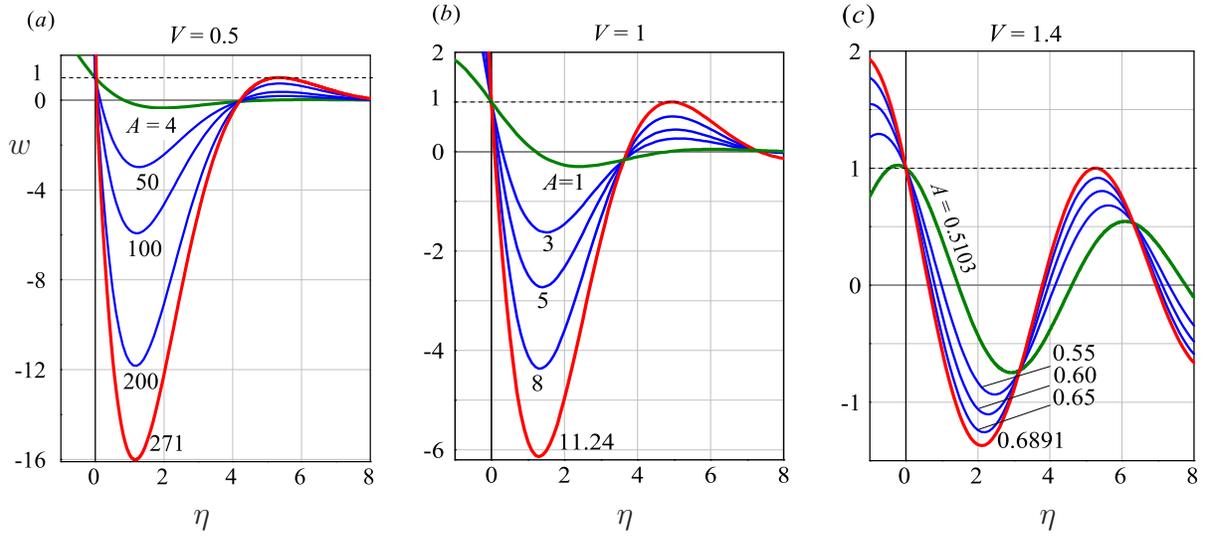}\hspace{5mm}
    \begin{picture}(0,0)(-100,-120)
\put(-90,80){$w$}
\put(140,-60){$\eta$}
\put(-20,-60){$\eta$}
\put(300,-60){$\eta$}
\end{picture}

\vspace{-20mm}
\caption{The steady-state regime. Displacements, $w(\Gn)$ (mainly at $\Gn>0$) corresponding to the incident wave phase speeds $V=0.5, 1, 1.4$ and to different amplitudes beginning from the lower bounds \eq{tlbfsss} (green in electronic version) and until the upper bounds (red in electronic version), where the first maximum is equal to $w_c$ ($A$ is the incident wave amplitude). }
\label{f3}
\end{figure}

\begin{figure}[h!]

\vspace{-5mm}    \hspace{35mm}\includegraphics [scale=1.5]{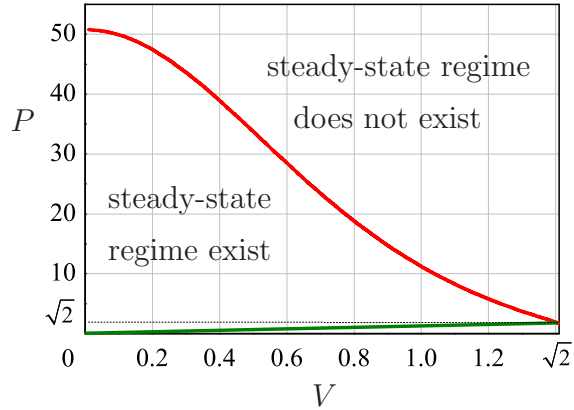}\hspace{5mm}
\begin{picture}(0,0)(200,-100)
\put(-20,-10){steady-state }
\put(-20,-30){regime exist}
\put(40,40){steady-state regime}
\put(50,20){does not exist}
\put(-57,19){$P$}
\put(57,-85){$V$}
\end{picture}

\vspace{-5mm}
\caption{The lower (green in electronic version) and upper (red in electronic version) bounds of the domain, where the steady-state solution does exist ($P=V^3A$; $A$ and $V$ are the incident wave amplitude and phase speed, \res). }
\label{f4}
\end{figure}

The corresponding plots for $V= 0.5, 1, 1.4$ are presented in \fig{f3}, where the results for different incident wave amplitudes are shown beginning from the lower bounds \eq{tlbfsss} and until the upper bounds, where the first maximum is equal to $w_c$. The plots evidence that the steady-state regime exists in the domain between these bounds and does not exist outside it. The bounds plotted based on the analytical solution are shown in \fig{f4}.

\subsection{Forerunning mode transition}

\begin{figure}[t]

\vspace{-45mm}    \hspace{-0mm}\includegraphics [scale=1.6]{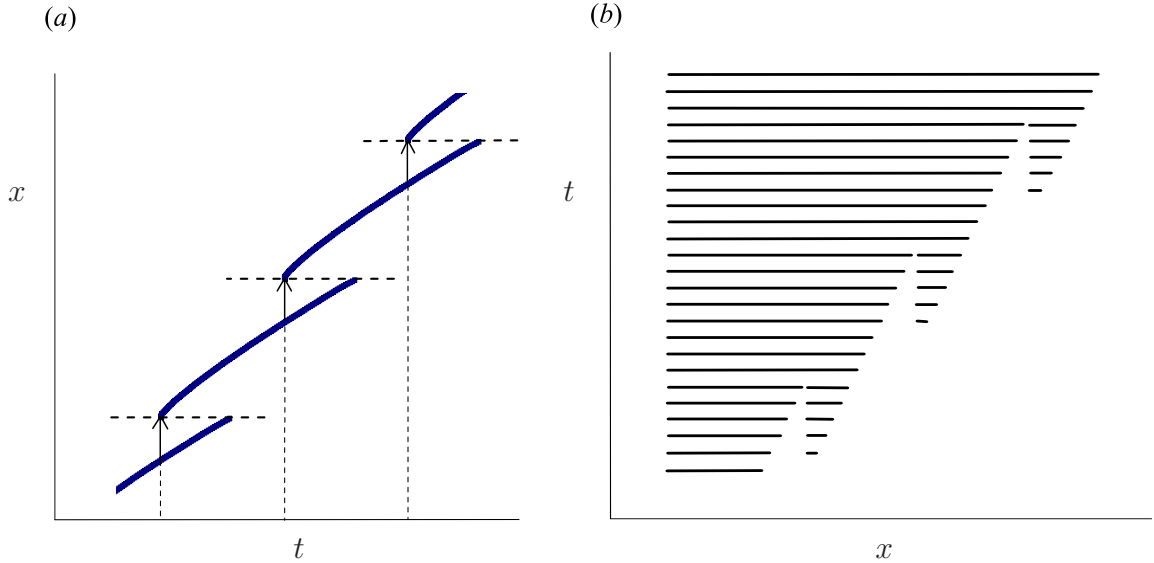}\hspace{5mm}
\begin{picture}(0,0)(0,-200)
\put(22,0){$x$}
\put(130,-135){$t$}
\put(232,0){$t$}
\put(350,-135){$x$}
\end{picture}

\vspace{-20mm}
\caption{Two different schemes of the forerunning mode transition: (a) A piecewise continuous curve as a two-valued function, $x(t)$. Origination of the forerunner in front of the main transition wave is indicated by arrows placed at $t_n$, $t_n\pm T, ...$, where $T$ is the period. Note that the end points corresponding to the same $x$ represent a single point reached by the main transition wave and a forerunner at different moments. At the moment, when the main transition wave reaches this point the beam-foundation separation line gains the forerunner-length increment. (b) Scheme of the beam-foundation separation lines represented for continuous $x$ and discrete values of $t$. The main transition wave and periodically originating and developing forerunners are shown. }
\label{f4b}
\end{figure}

What happens when the incident wave amplitude appears on the upper boundary, that is, when the displacement reaches the critical value at $\Gn>0$. The analytical solution suggests in this case that the  forerunner transition occurs at a distance of the main transition wave in front of it, and the steady-state mode of the transition fails. The established forerunning mode studied numerically appears periodic; its scheme is presented in \fig{f4b}. We present two graphical schemes of this nontrivial beam-foundation separation mode. In one of them, the separation path, $x(t)$, is a two-valued function of continuous time, \fig{f4b}a, whereas snapshots of the corresponding lines at a discrete set of time are shown in the other, \fig{f4b}b.

\begin{figure}[t]

\vspace{-105mm}    \hspace{-70mm}\includegraphics [scale=3.5]{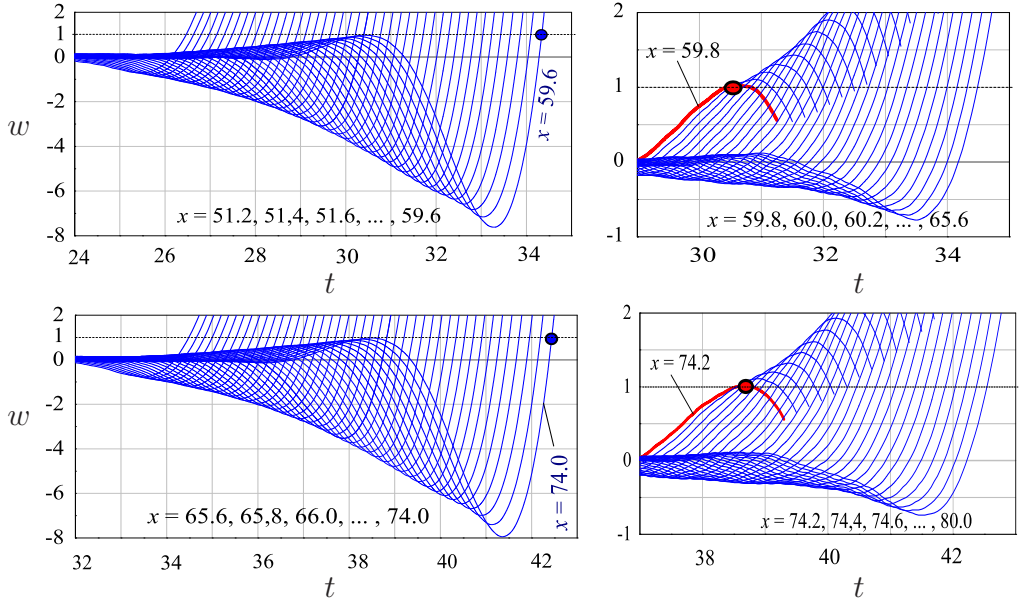}\hspace{5mm}
\begin{picture}(0,0)(0,-180)
\put(30,60){$w$}
\put(30,-50){$w$}
\put(150,0){$t$}
\put(350,0){$t$}
\put(150,-115){$t$}
\put(350,-115){$t$}
\end{picture}

\vspace{-20mm}
\caption{The forerunning mode transition. The function, $w(x,t)$, for a set of $x$-points is plotted for $V=1$. In an initial stage shown in the left plots, a regular transition wave propagates, and the evanescent wave penetrated into the intact area remains below the critical level. However, it increases, at a moment it reaches the threshold (the bold curve in the right plots (red in the electronic version)), and the forerunner transition arises. }
\label{f7}
\end{figure}

\vspace{5mm}
The established forerunning mode obtained in the numerical simulations is illustrated by \fig{f7}. In an initial stage shown in the plot, a regular transition wave propagates, and the evanescent wave penetrated into the intact area remains below the critical level. However, the latter increases gradually in time, and at a moment it reaches the threshold giving rise to the forerunner. This scenario is repeated periodically. Note that in the representation of the numerical simulation results, we normalise the displacement and the incident wave amplitude, $w(x,t)$ and $A$, attributed it to $w_c$. So we write $w(x,t)$, $A$ and 1 instead of $w(x,t)/w_c$, $A/w_c$ and $w_c$, \res.

\begin{figure}[h!]

\vspace{-80mm}    \hspace{0mm}\includegraphics [scale=3.3]{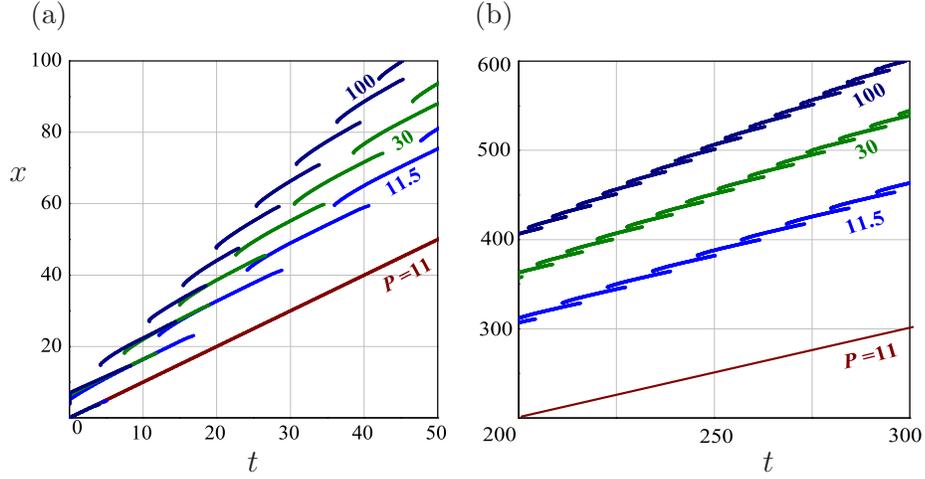}\hspace{5mm}
\begin{picture}(0,0)(0,-180)
\put(83,70){\small (a)}
\put(251,70){\small (b)}
\put(75,10){$x$}
\put(165,-100){$t$}
\put(360,-100){$t$}
\end{picture}

\vspace{-25mm}
\caption{The development of the forerunning transition wave in the intersonic regime for $V=1$. An initial stage is shown in \fig{f5}s, and the established regime is demonstrated in \fig{f5}b. The lower lines correspond to the steady-state regime. Recall that for $V=1$ the latter fails at $P=A=11.24$. In these plots, the local and averaged over the period speeds can be estimated.}
\label{f5}
\end{figure}

The development of the forerunning transition wave in the intersonic regime ($V<\sqrt{2}$) can be observed in \fig{f5} for some values of the incident wave amplitudes, $A$ ($P=V^3A$). An initial stage is shown in \fig{f5}a, and the established regime is demonstrated in \fig{f5}b. The lower lines correspond to the steady-state regime. Recall that for $V=1$ the latter fails at $A=11.24$. In these plots, the local and averaged over the period speeds can be estimated.

\begin{figure}[h!]

\vspace{-15mm}    \hspace{-0mm}\includegraphics [scale=2.0]{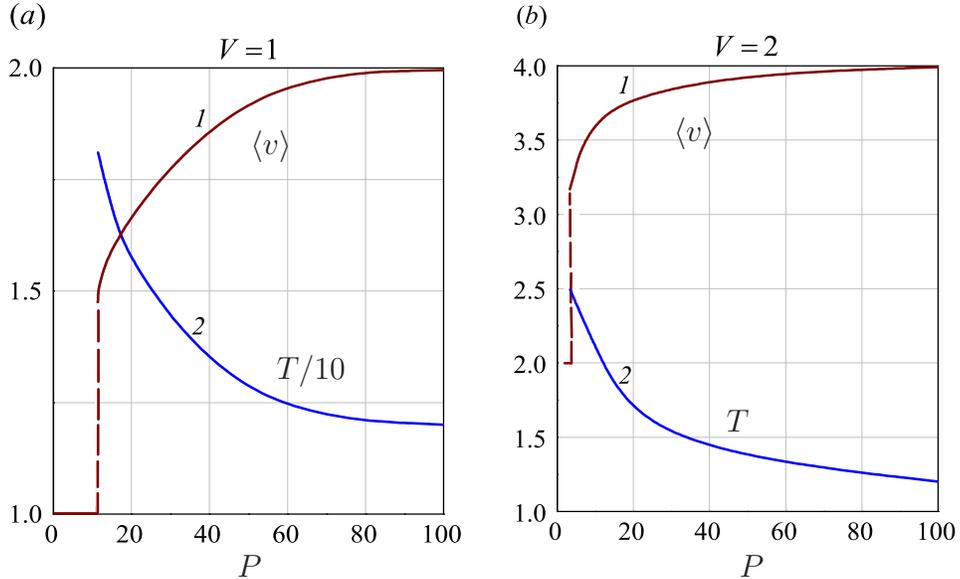}\hspace{5mm}
\begin{picture}(0,0)(0,-320)
\put(150,60){$\langle v \rangle$}
\put(160,-25){$T/10$}
\put(310,65){$\langle v \rangle$}
\put(330,-45){$T$}
\put(145,-100){$P$}
\put(335,-100){$P$}
\end{picture}

\vspace{-70mm}
\caption{Dependencies of the transition wave averaged speed, $\langle v \rangle$ (curves 1),  and the forerunning mode period, $T$ (curves 2), on the wave amplitude, $A$ $(P=V^3A)$, for the intersonic regime (a) and supersonic regime (b).}
\label{f6}
\end{figure}

The graphs of the transition wave averaged speed and the forerunning mode period as functions of the wave amplitude are presented in \fig{f6} for the intersonic ($V<\sqrt{2}$) and supersonic ($V>\sqrt{2}$) regimes, \fig{f6}a and \fig{f6}b, \res.

\begin{figure}[h!]

\vspace{-0mm}    \includegraphics [scale=1.6]{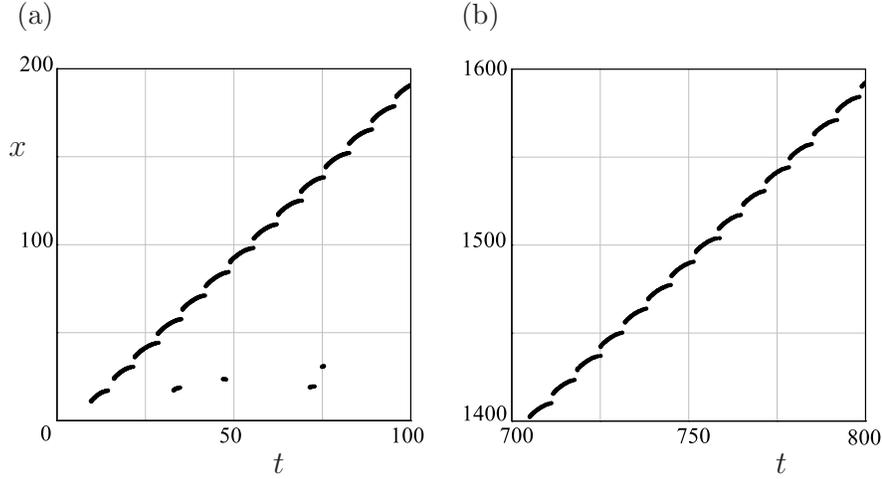}\hspace{5mm}
\begin{picture}(0,0)(450,-500)
\put(83,20){\small (a)}
\put(251,20){\small (b)}
\put(80,-30){$x$}
\put(180,-150){$t$}
\put(370,-150){$t$}
\end{picture}

\vspace{-120mm}
\caption{The `bridged-crack' periodic transition mode in the supersonic regime for different time intervals, $V=2$, $P=V^3A=0.8$. Separate points shown in \fig{f9}a correspond to local damages of some bridges detected only in the initial region of the transition.
The other bridges remain intact in the calculation period, $0< t < 800$.}
\label{f9}
\end{figure}

\begin{figure}[h!]

\vspace{-0mm} \hspace{10mm}   \includegraphics [scale=1.3]{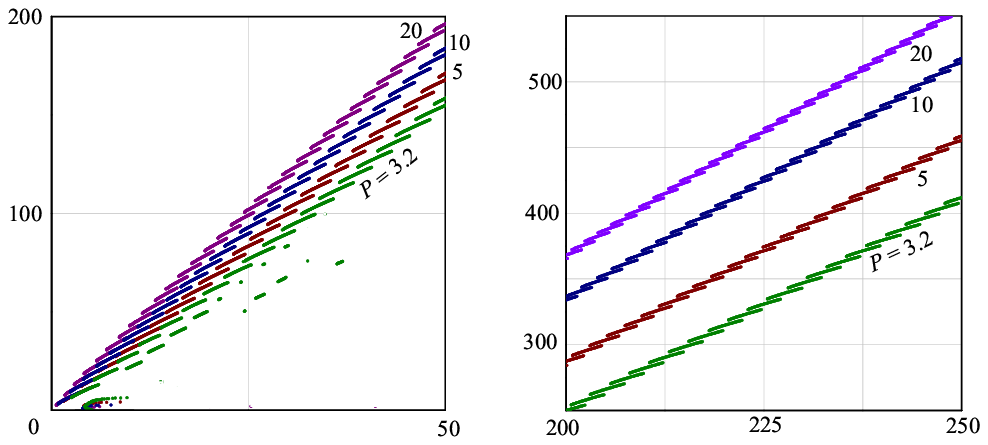}\hspace{5mm}
\begin{picture}(0,0)(0,-100)
\put(60,130){\small (a)}
\put(255,130){\small (b)}
\put(55,65){$x$}
\put(250,65){$x$}
\put(180,-50){$t$}
\put(370,-50){$t$}
\end{picture}

\vspace{-10mm}
\caption{The development of the forerunning transition wave in the supersonic regime for $V=2$, $P=V^3A=3.2, 5, 10, 20$ shown in an initial time interval (a) and for the established regime (b).}
\label{f10}
\end{figure}

In the supersonic regime, the calculations were performed for $V=2$. Established periodic transition modes were found in a rather narrow vicinity of $P=0.8$, \fig{f9} and for $P\ge 3.2$, \fig{f10}. The former is similar to a bridged crack, where the forerunners are not merged, whereas in the latter range, the forerunning mode appears similar to that detected in the intersonic regime.

\section{Discussions and conclusions}
A new transition wave is found propagating in the continuous waveguide under the action of a sinusoidal wave. It represents the periodic forerunning mode transition, which replaces the steady-state mode. The domain in $(V,A)$-plane is determined where the steady-state regime does exist. The forerunning mode transition exists outside this domain.

This transition mode is described in detail.
The relationships between the forerunning mode period and speed averaged over the period on the incident wave amplitude and speed are determined. The mechanisms of the transformation of the steady-state mode transition into the forerunning one and of the development of the latter are elucidated. In particular, it is shown that the forerunning mode transition speed increases to the group speed of the incident wave as intensity of the latter increases. This is in contrast to the steady-state regime, where the transition and incident wave speeds coincide independently of the intensity of the incident wave.

Our considerations are based on the  Euler-Bernoulli equation and Winkler foundation models. However, the phenomenon discovered in this paper has a more general nature. Below we describe the set of the necessary conditions at which the steady-state and forerunning transition waves can exist.

First of all, if an energy release rate is required for the transition (in the considered case, it is the critical energy of the foundation, $w_c^2/2$) the latter can propagate uniformly if

\begin{itemize}
\item[(a)]
{\it The group speed of the incident wave exceeds the phase speed}. This condition (it is referred to the anomalous dispersion) follows from the fact that, in the steady-state regime, the transition speed coincides with the incident wave phase speed, whereas the energy flux speed is equal to the group speed.

\item[(b)]
{\it No sinusoidal waves of the same phase speed must be in the intact region}. If this condition is satisfied and the steady-state regime exists, there is the total internal reflection from the moving transition front, and only an evanescent wave(s) penetrates into the intact region ahead of the transition front. Otherwise, Marder-Gross admissibility condition does not hold, and the forerunning mode transition occurs instead of the steady-state regime.
\item[(c)]
{\it The evanescent wave oscillates or at least is not monotonically decreasing}. (Note that this may follow from the condition (a).) In this case, the steady-state regime is replaced by the forerunning transition mode as the incident intensity reaches a threshold. Indeed, since the transition and incident wave speeds coincide and the transition criterion is fixed, the transition front changes its position relatively to the incident wave moving `downhill' as intensity of the latter grows (see \eq{ssds2} and \eq{ssds6}). As the result, the maximum of the evanescent wave can grow and reach the transition criterion at a distance of the `main' transition front as is demonstrated in this paper.
\end{itemize}

Finally we note that Euler-Bernoulli equation, in its application to an elastic beam, is valid if the wave length, $2\pi/k$, is much greater than the beam cross-section height. In terms of the dimensional parameters, $k=k_d$ and $r$, with refer to \eq{nndu1} this reads
 \beq k_dr=k\gl(\f{\Gk r^2}{EA}\gr)^{1/4}\ll 1\,,\eeq{ebwc}
where for the intersonic regime the non-dimensional wavenumber $|k|=V < \sqrt{2}$.

\vspace{3mm}
{\bf Acknowledgements.} The work is supported by the FP7 PEOPLE Marie Curie IAPP grant
 No. 284544-PARM2.

\vspace{10mm}
\vskip 18pt
\begin{center}
{\bf  References}
\end{center}
\vskip 3pt

\inh Ayzenberg-Stepanenko, M.V., Mishuris, G.S., and Slepyan, L.I., 2014. Brittle fracture in a periodic structure with internal potential energy. Spontaneous crack propagation.  Proc. R. Soc. A 2014 470, 20140121.

\inh Brun, M., Movchan, A., B., and Slepyan, L. I., 2013. Transition wave in a supported heavy beam. J. Mech. Phys. Solids 61 (10),  2067–2085.
DOI: http://dx.doi.org/10.1016/ j.jmps.2013.05.004 .

\inh Cai, C.W., Cheung, Y.K., Chan, H.C., 1988.  Dynamic response of infinite continuous beams subjected to a moving force—An exact method.
Journal of Sound and Vibration  123 (3), 461-472.

\inh Friba, L., 1999.  Vibration of solids and structures under moving loads (Third addition). ASCE Press, USA.

\inh Marder, M., and Gross, S., 1995. Origin of crack tip instabilities. J of the Mech. Phys. Solids 43, 1 - 48.

\inh Mishuris, G.S., Movchan, A.B., Slepyan, L.I., 2009. Localised knife
waves in a structured interface. J. Mech. Phys. Solids 57, 1958-1979.
DOI: 10.1016/j.jmps.2009.08.004

\inh Slepyan, L.I., 1981a. Dynamics of a crack in a lattice. Sov.
Phys. Dokl., 26, 538-540.

\inh Slepyan, L.I., 1981b. Crack Propagation in High-frequency
Lattice Vibration. Sov. Phys. Dokl., 26, 900-902.

\inh Slepyan, L.I., 2002. Models and Phenomena in Fracture
Mechanics. Springer, Berlin.

\inh Slepyan, L.I., and Ayzenberg-Stepanenko, M.V., 2004. Localized
transition waves in bistable-bond lattices. J. Mech. Phys. Solids
52(7), 1447-1479.

\inh Slepyan, L.I., 2010. Dynamic crack growth under Rayleigh wave. J. Mech. Phys. Solids 58, 636-655. doi:10.1016/j.jmps.2010.03.003
http://dx.doi.org/10.1016/j.jmps.2010.03.003

\inh Slepyan, L.I., Mishuris, G.S., Movchan, A.B., 2010.  Crack in a lattice waveguide. Int J Fract 162, 91-106.
DOI: 10.1007/s10704-009-9389-5

\inh Vainchtein, A., 2010. The role of spinodal region in the kinetics of lattice phase transitions. J. Mech. Phys. Solids, 58(2): 227-240.

\inh Cai, C.W., Cheung, Y.K., Chan, H.C., 1988.  Dynamic response of infinite continuous beams subjected to a moving force—An exact method.
Journal of Sound and Vibration  123 (3), 461-472.

\end{document}